\documentclass[epj]{svjour}
\usepackage{graphicx}
\usepackage{epstopdf}
\usepackage{amsmath}
\usepackage{amsfonts}
\usepackage{amssymb}
\usepackage{color}
\usepackage{mathrsfs}
\usepackage{multirow}
\usepackage{bm}
\usepackage{subfigure}
\usepackage{xcolor}
\usepackage{mathtext,bbm,amsmath,amsfonts,amssymb,indentfirst,syntonly,graphicx}
\usepackage{mathtools}
\usepackage{slashbox}
\usepackage[english]{babel}
\usepackage{calc}
\usepackage{tikz}
\usepackage[colorlinks,
linkcolor=blue,
anchorcolor=blue,
citecolor=blue
]{hyperref}
\usepackage{cite}
\usepackage{threeparttable}
\begin{document}
\title{Constraining the anisotropy of the Universe with the X-ray and UV fluxes of quasars}
\author{Dong Zhao, Jun-Qing Xia
\thanks{xiajq@bnu.edu.cn}%
}                     
%
%
\institute{Department of Astronomy, Beijing Normal University, Beijing 100875, China}
\date{Received: date / Revised version: date}
%
\abstract{
We test the anisotropy in the Finslerian cosmological model with the X-ray and ultraviolet (UV) fluxes of 808 quasars. The dipole amplitude is $A_D=0.302_{ -0.124}^{ +0.185}$ and the dipole direction points towards $(l, b) = ( 288.92_{~ -28.80^{\circ}}^{^{\circ}+23.74^{\circ}}, 6.10_{~ -16.40^{\circ}}^{^{\circ} +16.55^{\circ}} )$. We find that the dipole direction from the X-ray and UV fluxes of quasars is very close to the dipole direction given by the ``Joint Light-curve Analysis" (JLA) compilation in the Finslerian cosmological model and the angular difference between the two dipole directions is only $10.44^{\circ}$. We also find the angular difference between the dipole direction from the 808 quasars in the Finslerian cosmological model and ones from the supernovae of type Ia (SNe Ia) samples in the dipole-modulated $\Lambda$CDM model is around $30^{\circ}$. Six gravitationally lensed quasars are considered to investigate the Hubble constant $H_0$ in the Finslerian cosmological model. We get a slightly smaller $H_0$ than the result given by the six gravitationally lensed quasars. Finally, we forecast the future constraints on the dipole parameters with the X-ray and UV fluxes of quasars. As the number of simulations increases, the precisions of the parameters related to anisotropy in the Finslerian cosmological model improve significantly. The X-ray and UV fluxes of quasars have a promising future as a probe of anisotropy in Finsler spacetime.
} 
\maketitle
\section{Introduction}\label{sec:introduction}
The Universe is homogeneous and isotropic on large scale, which is called the cosmological principle. Based on it, the standard Lambda cold dark matter ($\Lambda$CDM) cosmological model has been established. In the past few decades, many experiments test its validity and verify that it is consistent with most cosmological observations. The observations of Cosmic Microwave Background (CMB) temperature anisotropies and polarizations from Wilkinson Microwave Anisotropy Probe (WMAP) \cite{Hinshaw:2012aka,Bennett:2012zja} and Planck satellites \cite{Akrami:2018vks,Aghanim:2019ame,Aghanim:2018eyx,Akrami:2019bkn} provide high-precision constraints on the six based cosmological parameters. However, there still exist several anomalies that have been reported, such as the alignment of low-$\ell$ multipoles in the CMB temperature anisotropies \cite{Tegmark:2003ve,Bielewicz:2004en,Copi:2013jna,Chang:2013lxa}, the parity asymmetry \cite{Akrami:2019bkn,Kim:2010gd,Kim:2010gf,Gruppuso:2010nd,Liu:2013wfa,Zhao:2013jya} and the hemispherical power asymmetry \cite{Akrami:2019bkn,Eriksen:2003db,Hansen:2004vq,Eriksen:2007pc} in CMB, the spatial variation of the fine structure constant \cite{Webb:2010hc,King:2012id}, the anisotropic accelerating expansion of the Universe \cite{Bonvin:2006en,Cai:2013lja,Koivisto:2008ig,Chang:2014wpa,Chang:2014nca,Lin:2016jqp}, the alignment of quasar polarization vectors on large scale \cite{Hutsemekers:2005iz}, the MOND acceleration scale \cite{Zhou:2017lwy,Chang:2018vxs,Chang:2018lab}. These phenomena may imply that our Universe has a preferred direction.

As the most luminous and persistent energy source, quasars have extraordinary potential in the exploration of our Universe. In recent years, quasars are tentatively used to investigate the cosmological parameters. An incomplete list includes the relation between the UV emission lines and the continuum luminosity \cite{1977ApJ...214..679B}, the relation between the radius of quasars and its luminosity \cite{Watson:2011um,Melia:2013sxa,Kilerci_Eser_2015}, the relation between luminosity and mass of super Eddington accreting quasars \cite{Wang:2013ha}, the correlation between X-ray variability and luminosity of quasars \cite{LaFranca:2014eba}, the non-linear relation between UV and X-ray luminosity \cite{Risaliti:2015zla,Khadka:2019njj,Khadka:2020tlm,Lusso:2020pdb}. The non-linear relation between UV and X-ray luminosity was firstly discovered by the X-ray surveys \cite{1979ApJ...234L...9T,1981ApJ...245..357Z,1986ApJ...305...83A}. For decades, the UV and X-ray luminosity relationship has been confirmed by observations of a few hundred quasars in the redshift range from 0 to 6.5. Since 2015, E. Lusso etc. \cite{Risaliti:2015zla,Lusso:2020pdb} have been attempting to estimate the cosmological parameters by the non-linear relation between UV and X-ray luminosity with quasars as standardizable candles. 

In this paper, we will use the X-ray and UV fluxes of 808 quasars \cite{Risaliti:2015zla} to explore the anisotropy in the Universe. These 808 quasars are thought to be standardizable candles through the relation between UV and X-ray luminosity. The quasars sample is in the redshift range $0.061\leq z \leq 6.28$. The Pantheon sample \cite{Scolnic:2017caz} and the JLA compilation \cite{Betoule:2014frx} are combined with 808 quasars in the analysis of the anisotropic cosmological model i.e., Finslerian cosmological model. We also attempt to investigate the Hubble constant $H_0$ in the Finslerian cosmological model by considering six gravitationally lensed quasars with measured time delays \cite{Wong:2019kwg}. At last, we will forecast the future constraints on the Finslerian cosmological model with the X-ray and UV fluxes of quasars.


The rest of this paper is organized as follows. In Section \ref{sec:Methodology}, we briefly introduce the UV and X-ray luminosity relationship, the Time-Delay Strong Lensing measurement, and the Finslerian cosmological model. We show our results in Section \ref{sec:PJq}. Finally, discussions and conclusions are given in Section \ref{sec:DC}.
\section{Methodology}\label{sec:Methodology}
\subsection{The UV and X-ray luminosity relationship}
The relation of the UV and X-ray luminosity is parameterized as
\begin{equation}\label{lux_lx_a}
\alpha_{\mathrm{OX}}=0.384 \times \log \left(L_{\mathrm{X}} / L_{\mathrm{UV}}\right),
\end{equation}
where $L_{\mathrm{UV}}$ denotes the logarithm of the monochromatic luminosity at 2500 {\r A} and $L_{\mathrm{X}}$ denotes the logarithm of the monochromatic luminosity at 2 keV. $\alpha_{\mathrm{OX}}$ is the slope of power law, which connects $L_{\mathrm{X}}$ and $L_{\mathrm{UV}}$. Eq. (\ref{lux_lx_a}) can be written as
\begin{equation}\label{lux_lx}
\log L_{\mathrm{X}}=\beta+\gamma \log L_{\mathrm{UV}},
\end{equation}
where $\beta$ and $\gamma$ are two free parameters. The luminosities and fluxes of quasars are connected by the luminosity distance. Now we rewrite the Eq. (\ref{lux_lx}) and obtain
\begin{equation}\label{lux_lx_F}
\log \left(F_{X}\right)=\beta+(\gamma-1) \log (4 \pi)+\gamma \log \left(F_{U V}\right)+2(\gamma-1) \log \left(D_{L}\right),
\end{equation}
where $\log$ denotes $log_{10}$. $F_{X}$ and $F_{U V}$ represent the are X-ray and UV fluxes of quasars, respectively. The luminosity distance $D_{L}$ takes the form,
\begin{equation}\label{DL}
D_{L}=\frac{c\left(1+z\right)}{H_0} \int_{0}^{z} \frac{d z'}{E(z')},
\end{equation}
where $z_{c m b}$ denotes redshift. c is the speed of light and $H_0$ is the Hubble constant. The expression of $E(z)$ depends on cosmological models. 

\begin{figure}
	\begin{center}
		\includegraphics[width=9cm]{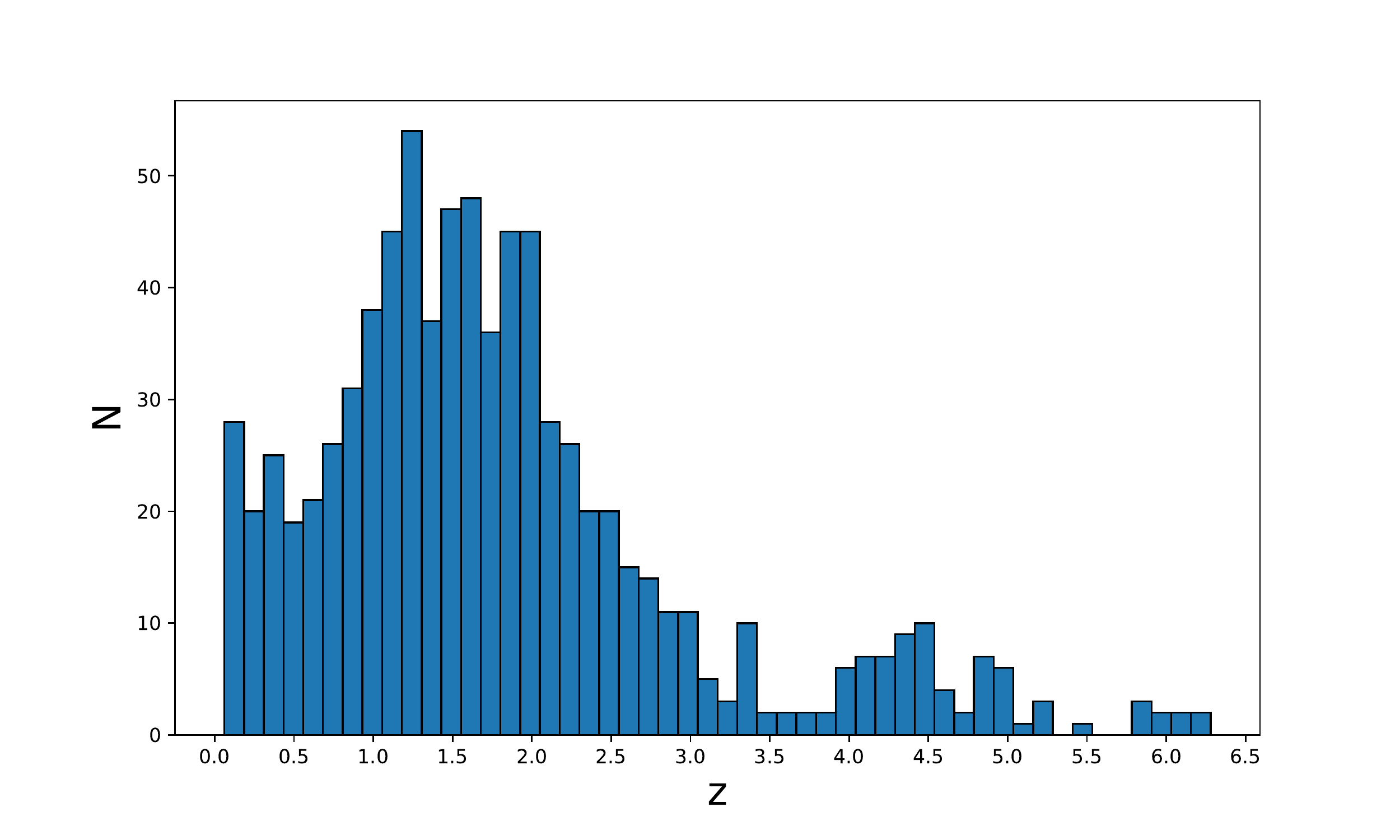}
		\caption{The redshift distribution of 808 quasars.}
		\label{fig:red_dis}
	\end{center}
\end{figure}

\begin{figure}
	\begin{center}
		\includegraphics[width=8.5cm]{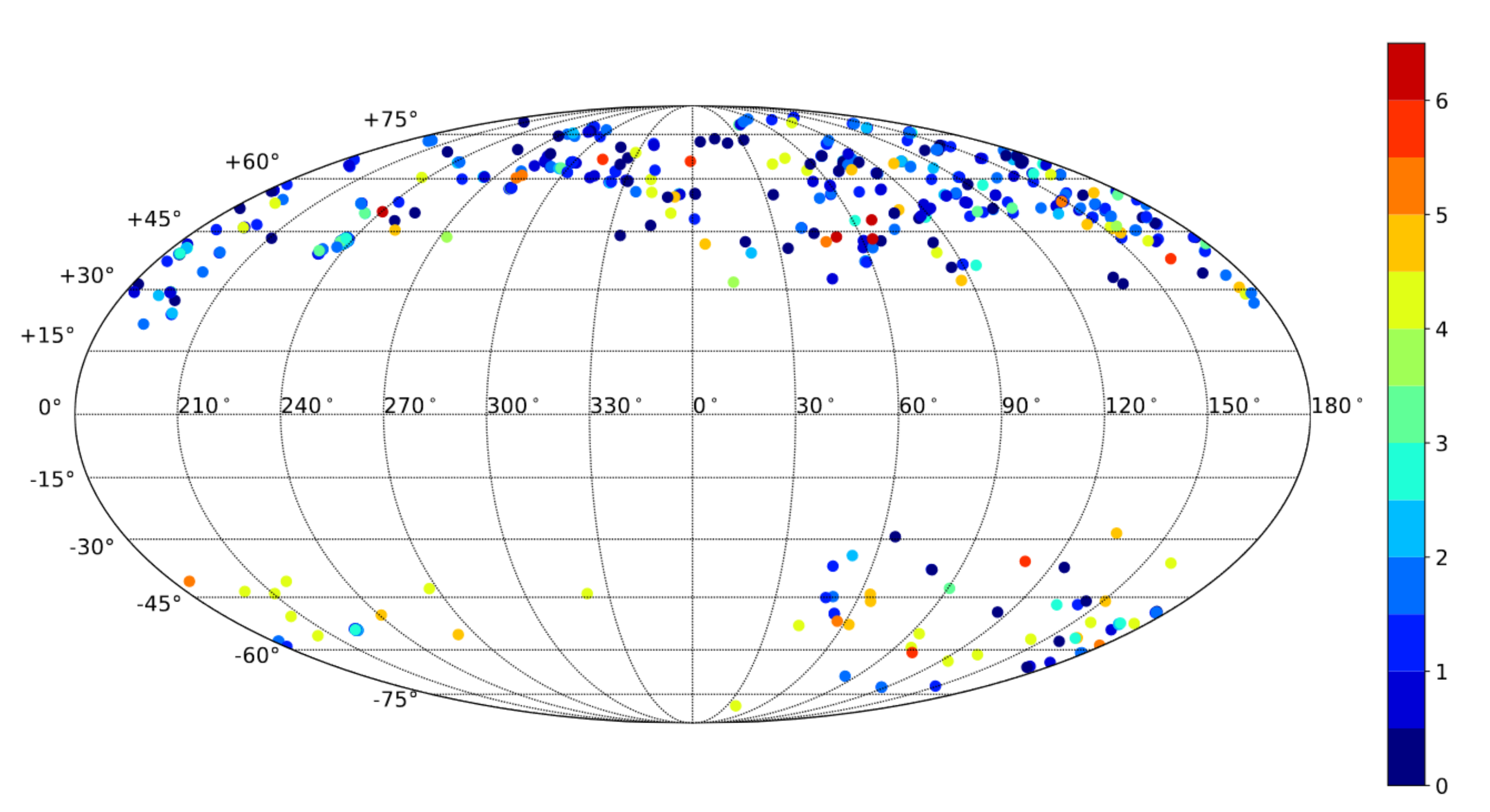}
		\caption{The distribution of 808 quasars in the galactic coordinate system. The pseudo-colors indicate redshift of these quasars.}
		\label{fig:red_dis_g}
	\end{center}
\end{figure}

In our work, the dataset of the X-ray and UV fluxes of quasars are from G. Risaliti and E. Lusso \cite{Risaliti:2015zla}. The dataset contains 808 quasars, which are in the redshift range $0.061\leq z \leq 6.28$. The redshift distribution of 808 quasars is shown in Fig. \ref{fig:red_dis}. The distribution of 808 quasars in the galactic coordinate system is shown in Fig. \ref{fig:red_dis_g} and the pseudo-colors indicate the redshift of these quasars.

\subsection{The Time-Delay Strong Lensing measurement}
Strong gravitational lensing is a powerful probe of cosmological models. The time-delay strong lensing (TDSL) measurement is a fully independent method to measure the Hubble constant.  Since the approach first proposed by Refsdal \cite{Refsdal:1964nw}, lensed quasars have generally been used to constrain $H_0$ by measuring the difference in arrival time of photons. TDSL provides a measurement of $H_0$, which is completely independent of the CMB and the local distance ladder.

The travel time of light rays from a source to the observer depends on their path length and the gravitational potential they traverse. For a system of lenses with an image at an angular position $\boldsymbol{\theta}$ and corresponding source position $\boldsymbol{\beta}$, the excess time delay is
\begin{equation}\label{e_time_delay}
t(\boldsymbol{\theta}, \boldsymbol{\beta})=\frac{D_{\Delta t}}{c}\left[\frac{(\boldsymbol{\theta}-\boldsymbol{\beta})^{2}}{2}-\psi(\boldsymbol{\theta})\right],
\end{equation}
where c is the speed of light and $\psi(\boldsymbol{\theta})$ is the lens potential. The time-delay distance $D_{\Delta t}$ is defined as \cite{Refsdal:1964nw}
\begin{equation}
D_{\Delta t} = \left(1+z_{\mathrm{d}}\right) \frac{D_{\mathrm{d}} D_{\mathrm{s}}}{D_{\mathrm{ds}}},
\end{equation} 
where $z_{\mathrm{d}}$ denotes the lens redshift. $D_{\mathrm{d}}$ and $D_{\mathrm{s}}$ are the angular diameter distance from the observer to the lens and the angular diameter distance from the observer to the source, respectively. $D_{\mathrm{ds}}$ is the angular diameter distance from the lens to the source. The angular diameter distance is defined as
\begin{equation}\label{DA}
D_{A}=\frac{c}{H_0 \left( 1+z \right)} \int_{0}^{z} \frac{d z'}{E(z')},
\end{equation}
where $z$ is the redshift and $H_0$ is the Hubble constant. The expression of $E(z)$ depends on cosmological models. The difference of excess time delays between two images A and B is
\begin{equation}
\Delta t_{AB}=\frac{D_{\Delta t}}{c}\left[\frac{\left(\boldsymbol{\theta}_{A}-\boldsymbol{\beta}\right)^{2}}{2}-\psi\left(\boldsymbol{\theta}_{A}\right)-\frac{\left(\boldsymbol{\theta}_{B}-\boldsymbol{\beta}\right)^{2}}{2}+\psi\left(\boldsymbol{\theta}_{B}\right)\right],
\end{equation}
where $\boldsymbol{\theta}_{A}$ and $\boldsymbol{\theta}_{B}$ are the positions of image A and B, respectively.

We use six gravitationally lensed quasars with measured time delays from H0LiCOW collaboration \cite{Wong:2019kwg} to constrain the Hubble constant and other cosmological parameters. Our work is based on the $H_0$ inference code\footnote{https://github.com/shsuyu/H0LiCOW-public/tree/master/H0\_inference\_code.} provided by Kenneth C. Wong etc. \cite{Wong:2019kwg}.
\subsection{The anisotropic cosmological model}
In this paper, we choose the Finslerian cosmological model as the anisotropic cosmological model. Different from the standard cosmological model, the Finslerian cosmological model has an intrinsically preferred direction that breaks the isotropy of the Universe. Many works about investigating the anisotropy of the Universe are based on the Finsler spacetime. For instance, investigating the cosmic anisotropy with supernovae of type Ia (SNe Ia) samples by the hemisphere comparison HC method \cite{Chang:2014nca,Zhao:2019azy} and the dipole fitting \cite{Chang:2017bbi,Lin:2016jqp,Lin:2015rza,Chang:2014nca,Chang:2019utc}, explaining the parity asymmetry and power deficit in the Finsler spacetime \cite{Chang:2018bjg}, the unified description for dipoles of the fine-structure constant and SNe Ia Hubble diagram \cite{Li:2015uda}.

In the Finsler spacetime, the scale factor $a$ takes the form \cite{Li:2015uda},
\begin{equation}\label{a_F}
a=\left(1+A_{D} \cos \theta\right) /(1+z).
\end{equation}
$A_{D}$ is a parameter in the Finsler spacetime, which can be regarded as the dipole amplitude. When $A_{D}=0$, the Finslerian cosmological model reduces to the $\Lambda$CDM model. $\cos \theta$ is the angle between the position of quasars and the preferred direction in the Finsler spacetime. By Eq. (\ref{a_F}), the luminosity distance in the Finsler spacetime can be written as
\begin{equation}
D_{L}=\frac{c\left(1+z\right)}{H_0} \int_{0}^{z} \frac{d z'}{E(z')},
\end{equation}
where $E(z)$ in the Finsler spacetime takes the form of 
\begin{equation}\label{Ez}
E(z)=\sqrt{\Omega_{m 0}(1+z)^{3}(1 + A_{D} \cos \theta)^{-3}+1-\Omega_{m 0}}.
\end{equation}
Plugging Eq. (\ref{Ez}) into Eq. (\ref{DL}) and Eq. (\ref{DA}), we can get the form of the luminosity distance and angular diameter distance in the Finslerian cosmological model, respectively.
\section{Results}\label{sec:PJq}
To constrain the dipole amplitude and the preferred direction in the Finslerian cosmological model with the X-ray and UV fluxes of quasars, we employ the likelihood function \cite{Khadka:2019njj},
\begin{equation}
\ln (\mathrm{LF})=-\frac{1}{2} \sum_{i=1}^{N}\left[\frac{\left[\log \left(F_{X, i}^{\mathrm{obs}}\right)-\log \left(F_{X, i}^{\mathrm{th}}\right)\right]^{2}}{s_{i}^{2}}+\ln \left(2 \pi s_{i}^{2}\right)\right],
\end{equation}
where $s_{i}^{2}=\sigma_{i}^{2}+\delta^{2}$. $\sigma_{i}^{2}$ is the error of the observed flux $F_{X, i}^{\mathrm{obs}}$ and $\delta$ is the global intrinsic dispersion. $F_{X, i}^{\mathrm{th}}$ is the theoretical flux at the redshift $z_i$. 

The Markov chain Monte Carlo (MCMC) method has been used to explore the whole parameters space in our work. Emcee\footnote{https://emcee.readthedocs.io/en/stable/} \cite{ForemanMackey:2012ig} as the Affine Invariant Markov chain Monte Carlo Ensemble sampler is widely used to investigate the parameters in astrophysics and cosmology. During the fitting process, we find that the parameters $\beta$, $\gamma$, and $\delta$ are insensitive to the $\Lambda$CDM model and the Finslerian cosmological model. The results of the three parameters in the two cosmological models are almost the same. For the sake of brevity and clarity, we only show the parameters related to the Finslerian cosmological model.

The flat prior of each parameter in the Finslerian cosmological model is
\begin{equation}
\begin{aligned}
&\Omega_{\mathrm{m}} \sim [0,1], A_D \sim [0,1], l \sim [-180,180], b \sim [-90,90].
\end{aligned}
\end{equation}

\begin{figure}
	\begin{center}
		\includegraphics[width=8cm]{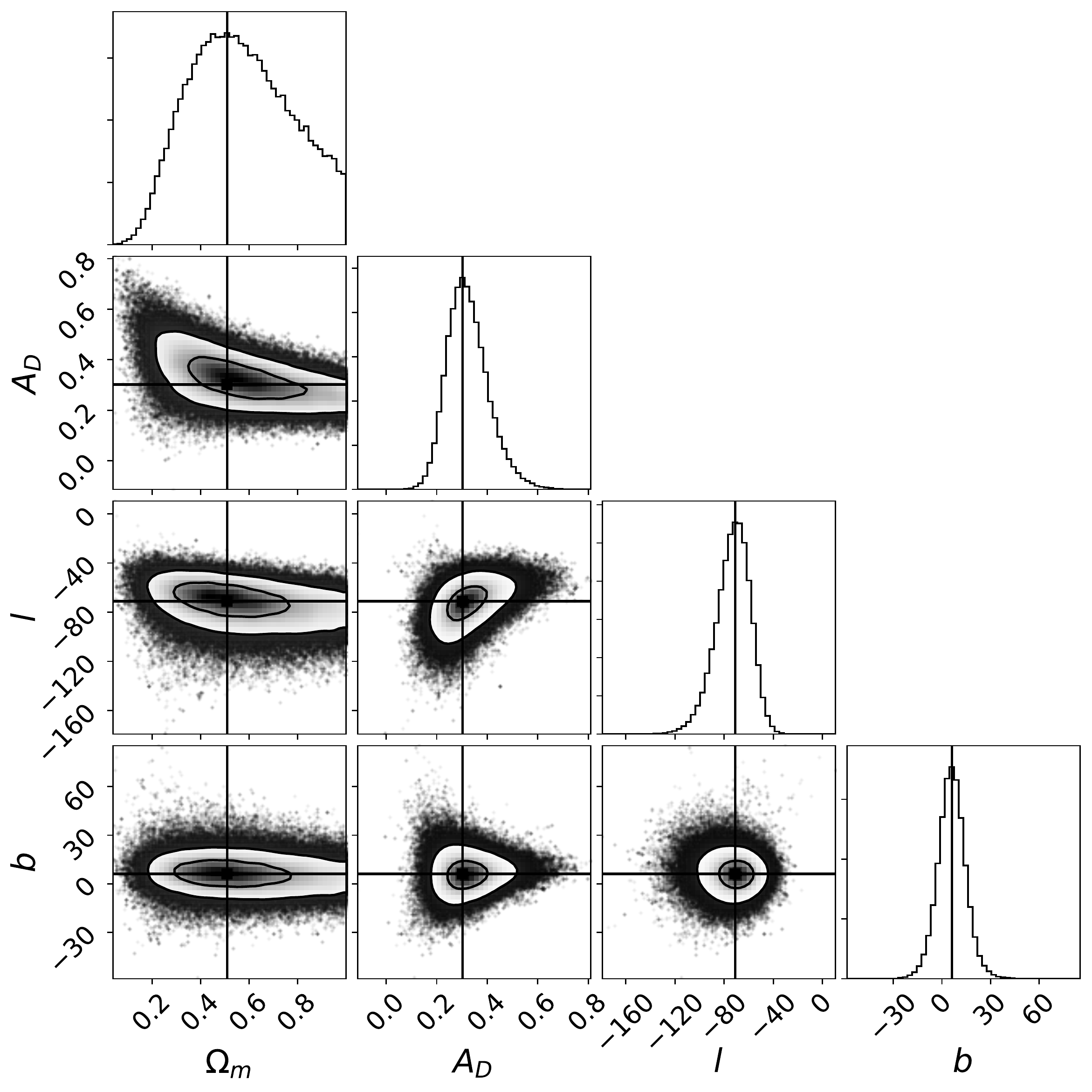}
		\caption{The marginalized posterior distribution of the parameters in the Finslerian cosmological model with the X-ray and UV fluxes of quasars. The horizontal and vertical solid black lines denote the maximum of 1-dimensional marginalized posteriors.}
		\label{fig:xuv_result}
	\end{center}
\end{figure}

The results are shown in Fig. \ref{fig:xuv_result} and summarized in Table \ref{table:C_data}. In Fig. \ref{fig:xuv_result}, we show the marginalized posterior distribution of the parameters. The horizontal and vertical solid black lines denote the maximum of 1-dimensional marginalized posteriors. In Table \ref{table:C_data}, we show the 68\% confidence level constraints on the parameters. As can be seen, the dipole anisotropy is well-constrained by the X-ray and UV fluxes of quasars. The dipole amplitude is $A_D=0.302_{ -0.124}^{ +0.185}$, which is not zero in 1$\sigma$ confidence level. The dipole direction points towards $(l, b) = ( 288.92_{~ -28.80^{\circ}}^{^{\circ}+23.74^{\circ}},$
$ 6.10_{~ -16.40^{\circ}}^{^{\circ} +16.55^{\circ}} )$. Compared with results from SNe Ia \cite{Lin:2015rza,Zhao:2019azy,Chang:2019utc}, the precision of dipole direction has a significant improvement. Interestingly, we found that the dipole direction from the X-ray and UV fluxes of quasars is very close to the dipole direction $(l, b) = ( 291.60_{~ -92.96^{\circ}}^{^{\circ} +248.10^{\circ}}, 16.20_{~ -78.73^{\circ}}^{^{\circ} +73.80^{\circ}} )$ given by the JLA in the Finslerian cosmological model. The angular difference between the two dipole directions is only $10.44^{\circ}$. The dipole direction given by the Pantheon sample in the Finslerian cosmological model is $(l, b) = ( 298.80_{~ -118.69^{\circ}}^{^{\circ} +75.31^{\circ}}, -23.41_{~ -57.41^{\circ}}^{^{\circ} +19.26^{\circ}} )$ \cite{Chang:2019utc}, which is about $31.05^{\circ}$ away from the dipole direction given by the X-ray and UV fluxes of quasars. 

\begin{table*}
	\begin{center}
	\caption{The 68\% confidence level constraints on the parameters in the Finslerian cosmological model with different datasets.}
	\begin{threeparttable}
	\setlength{\tabcolsep}{1.8mm}{
		\begin{tabular}{cccc}
			\hline Data & Quasars & Quasars + Pantheon + JLA & Quasars + TSDL \\
			\hline$\Omega_{\mathrm{m}}$ & $0.509_{ -0.275}^{ +0.453}$ & $ 0.298_{ -0.042}^{ +0.039}$ & $0.204_{ -0.137}^{ +0.190}$ \\
			$A_D$ & $0.302_{ -0.124}^{ +0.185}$ & $-$ & $0.142_{ -0.142}^{ +0.330}$ \\
			$l$ & $288.92_{~ -28.80^{\circ}}^{^{\circ}+23.74^{\circ}}$ & $284.41_{~ -104.37^{\circ}}^{^{\circ} +220.14^{\circ}}$ & $ 296.24_{~ -94.22^{\circ}}^{^{\circ} +46.62^{\circ}}$ \\
			$b$ & $6.10_{~ -16.40^{\circ}}^{^{\circ} +16.55^{\circ}}$ & $-9.00_{~ -80.99^{\circ}}^{^{\circ} +76.29^{\circ}}$ & $21.23_{~ -45.86^{\circ}}^{^{\circ} +51.33^{\circ}}$ \\
			$H_{0}$\tnote{1}  & $-$ & $-$ & $72.2_{-4.1}^{+3.6}$ \\
			\hline
	\end{tabular}}
\begin{tablenotes}
	\footnotesize
	\item[1] $\mathrm{~km} \mathrm{~s}^{-1} \mathrm{Mpc}^{-1}$
\end{tablenotes}
	\end{threeparttable}
	\label{table:C_data}
	\end{center}
\end{table*}

\begin{figure}
	\begin{center}
		\includegraphics[width=8cm]{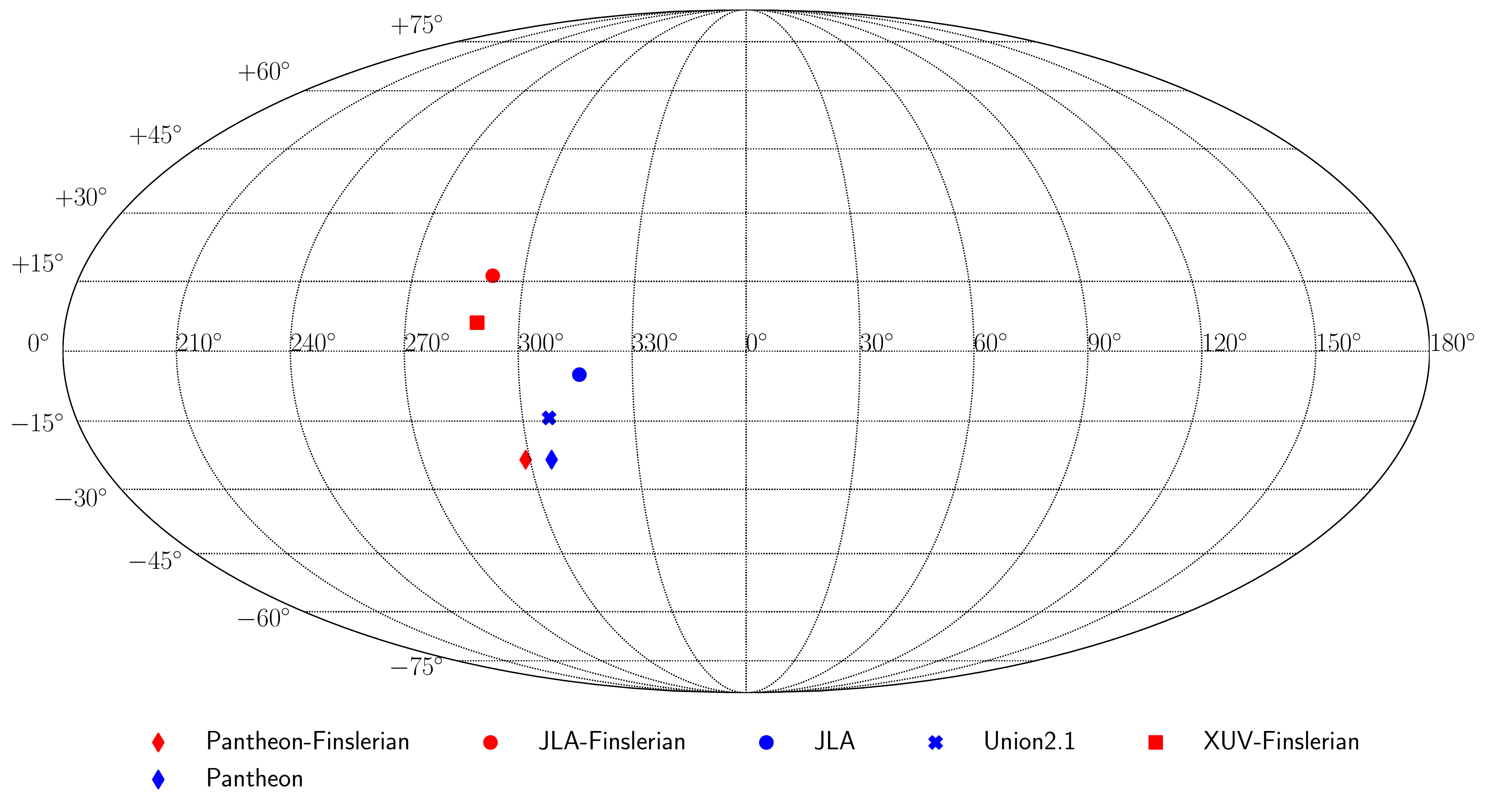}
		\caption{The dipole directions from the Finslerian cosmological model and the dipole-modulated $\Lambda$CDM model with different datasets in the galactic coordinate system.}
		\label{fig:DF-Compare}
	\end{center}
\end{figure}

The dipole directions from SNe Ia in the dipole-modulated $\Lambda$CDM model are also considered for comparison. In Table \ref{table:dipole_direction}, we summarized the dipole directions in the dipole-modulated $\Lambda$CDM model. In Fig. \ref{fig:DF-Compare}, we show all the dipole directions mentioned above in the galactic coordinate system. As can been seen in Fig. \ref{fig:DF-Compare}, the dipole direction given by the X-ray and UV fluxes of quasars is not far away from the dipole directions given by SNe Ia in the dipole-modulated $\Lambda$CDM model. We show the angular difference between the dipole direction given by the X-ray and UV fluxes of quasars and other dipole directions in Table \ref{table:difference}. For the three dipole directions in the dipole-modulated $\Lambda$CDM model, the angular difference is around $30^{\circ}$. The dipole direction from the X-ray and UV fluxes of quasars close to the ones from the SNe Ia sample, especially the JLA compilation, may hint that there could exist an underlying relation.

\begin{table}
	\caption{The 68\% confidence level constraints on the dipole directions in the dipole-modulated $\Lambda$CDM model with Pantheon \cite{Zhao:2019azy}, JLA \cite{Lin:2015rza}, and Union2.1 \cite{Yang:2013gea}.}
	\setlength{\tabcolsep}{1.8mm}{
		\begin{tabular}{cccc}
			\hline Data & Pantheon & JLA & Union2.1 \\
			\hline
			$l$ & $306.00_{~ -125.01^{\circ}}^{^{\circ} +82.95^{\circ}}$ & $316_{~ -110^{\circ}}^{^{\circ} +107^{\circ}}$ & $ 307.1^{\circ} \pm 16.2^{\circ}$ \\
			$b$ & $-34.20_{~ -54.93^{\circ}}^{^{\circ} +16.82^{\circ}}$ & $-5_{~ -60^{\circ}}^{^{\circ} +41^{\circ}}$ & $ -14.3^{\circ} \pm 10.1^{\circ}$ \\
			\hline
	\end{tabular}}
	\label{table:dipole_direction}
\end{table}

\begin{table*}
	\begin{center}
	\caption{The angular difference between the dipole direction given by the X-ray and UV fluxes of quasars and other dipole directions. ``F" denotes the Finslerian cosmological model and ``$\Lambda$" denotes the $\Lambda$CDM model.}
	\setlength{\tabcolsep}{1.8mm}{
		\begin{tabular}{cccccc}
			\hline Dipole direction & Pantheon-F & JLA-F  & Pantheon-$\Lambda$ & JLA-$\Lambda$ & Union2.1-$\Lambda$ \\
			\hline
			Difference & $31.05^{\circ}$ & $10.44^{\circ}$ & $33.90^{\circ}$ & $29.23^{\circ}$ & $27.23^{\circ}$\\
			\hline
	\end{tabular}}
	\label{table:difference}
	\end{center}
\end{table*}

\begin{figure}
	\begin{center}
		\includegraphics[width=8cm]{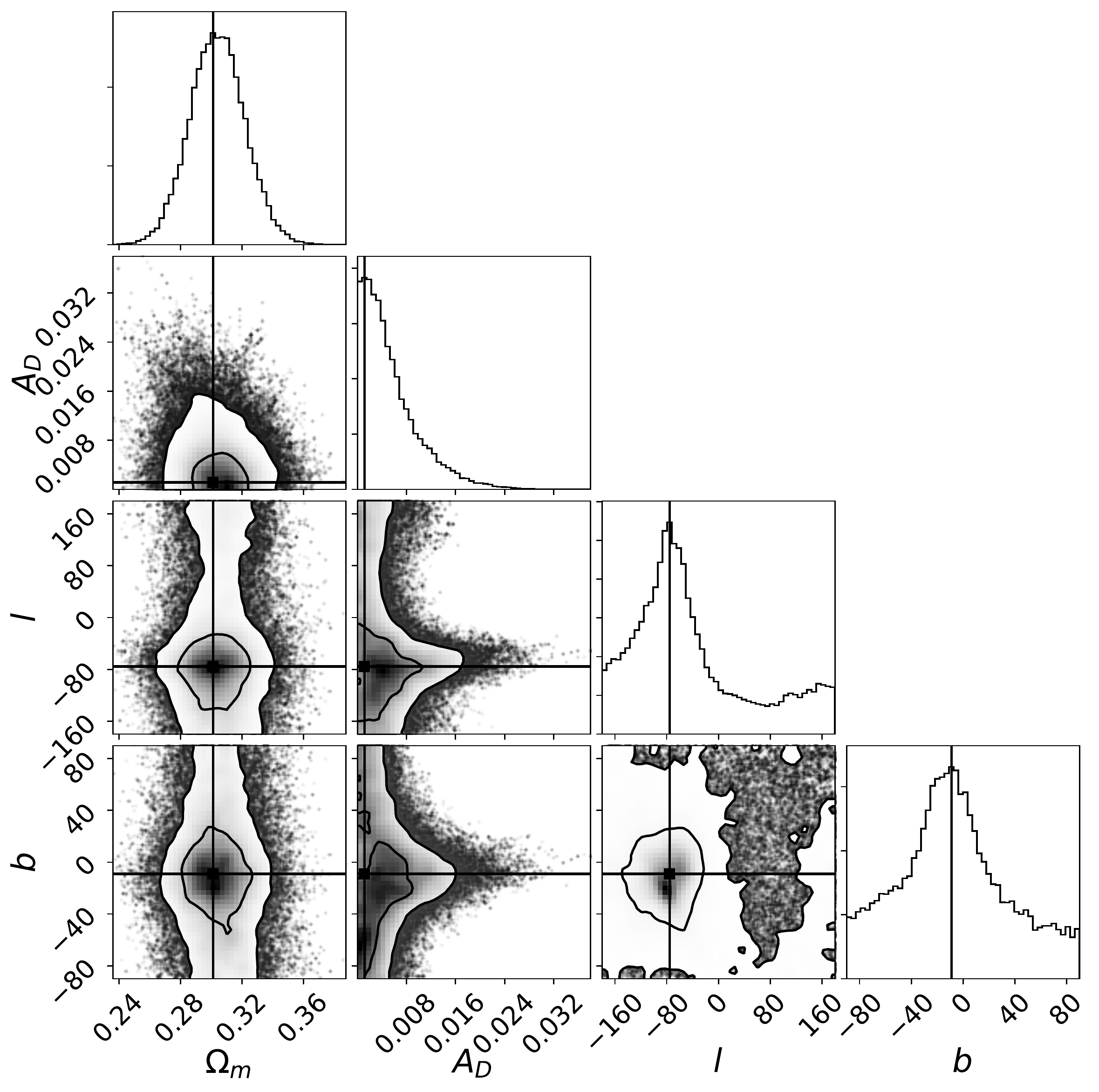}
		\caption{The marginalized posterior distribution of the parameters in the Finslerian cosmological model from quasars, Pantheon sample, and JLA compilation. The horizontal and vertical solid black lines denote the maximum of 1-dimensional marginalized posteriors.}
		\label{fig:PJX_result}
	\end{center}
\end{figure}

We combined the Pantheon sample and JLA compilation with quasars to constrain the Finslerian cosmological model. The results are shown in Fig. \ref{fig:PJX_result} and summarized in Table \ref{table:C_data}. In Fig. \ref{fig:PJX_result}, we show the marginalized posterior distribution of the parameters. The horizontal and vertical solid black lines denote the maximum of 1-dimensional marginalized posteriors. In Table \ref{table:C_data}, we show the 68\% confidence level constraints on the parameters. For the combined datasets, the 95\% confidence level upper limit of the dipole amplitude $A_D$ is $1.14\times 10^{-2}$ and the dipole direction points towards $(l, b) = ( 284.41_{~ -104.37^{\circ}}^{^{\circ} +220.14^{\circ}}, -9.00_{~ -80.99^{\circ}}^{^{\circ} +76.29^{\circ}} )$. The result is similar to one from the Pantheon sample in the Finslerian cosmological model \cite{Chang:2019utc}.






\begin{figure}
	\begin{center}
		\includegraphics[width=8cm]{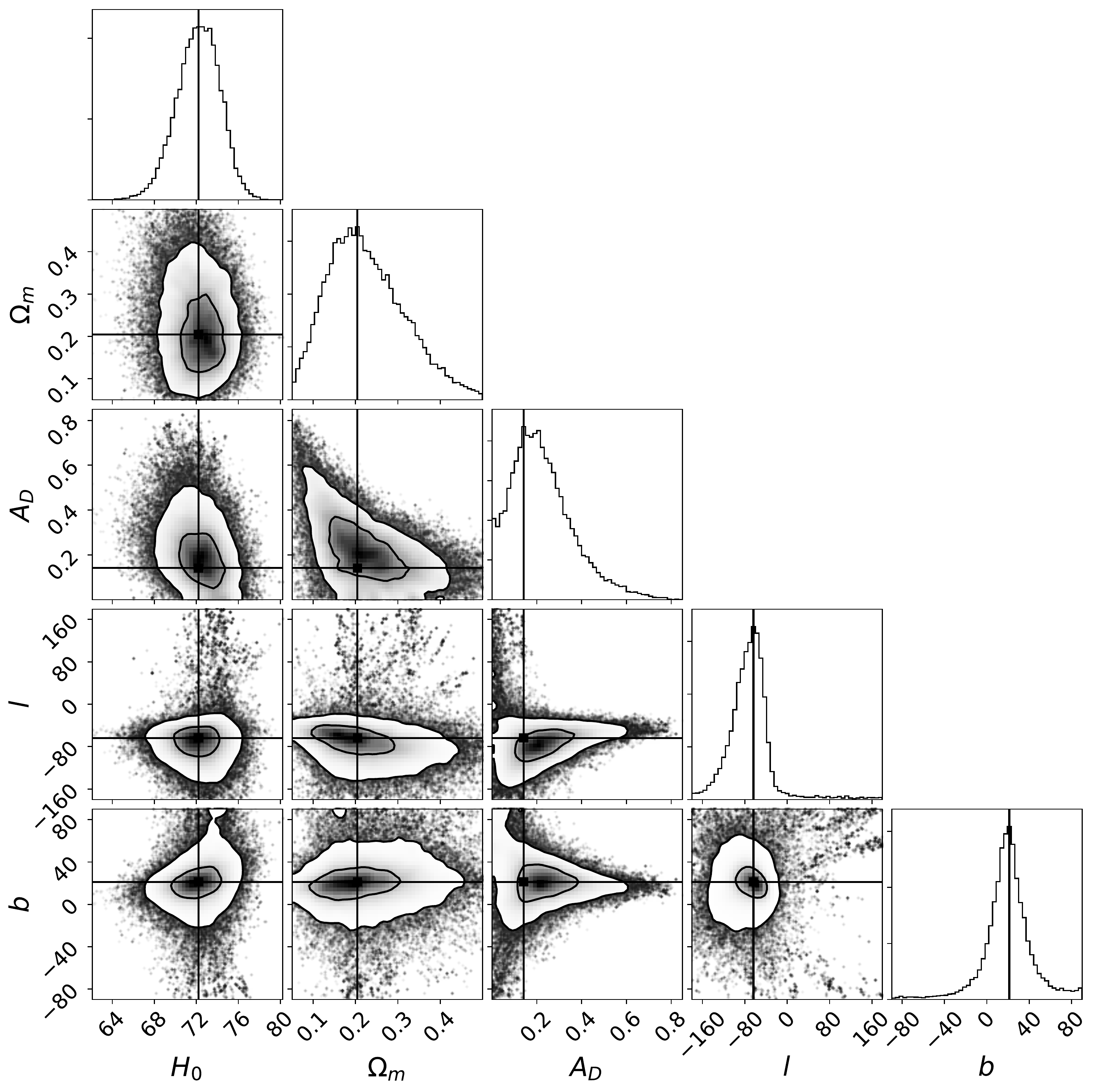}
		\caption{The marginalized posterior distribution of the parameters related to the Finslerian cosmological model with six gravitationally lensed quasars and the X-ray and UV fluxes of quasars. The horizontal and vertical solid black lines denote the maximum of 1-dimensional marginalized posteriors.}
		\label{fig:SLXUV_result}
	\end{center}
\end{figure}

To investigate the Hubble constant $H_0$ in the Finslerian cosmological model, we use six gravitationally lensed quasars with measured time delays from H0LiCOW collaboration \cite{Wong:2019kwg} in our analysis. The six gravitationally lensed quasars are combined with the X-ray and UV fluxes of quasars in the parameters fitting. The flat prior on each parameter is as follow: $H_0 \sim [60,100], \Omega_{\mathrm{m}} \sim [0,1], A_D \sim [0,1], l \sim [-180,180], b \sim [-90,90]$. We show the results in Fig. \ref{fig:SLXUV_result} and Table \ref{table:C_data}. We find that the dipole amplitude is $A_D=0.142_{ -0.142}^{ +0.330}$ and the dipole direction is $(l, b) = ( 296.24_{~ -94.22^{\circ}}^{^{\circ} +46.62^{\circ}}, 21.23_{~ -45.86^{\circ}}^{^{\circ} +51.33^{\circ}})$. Even though the errors of parameters are bigger than ones given by quasars, the results of parameters related to anisotropy are consistent with the results from quasars. The Hubble constant is $H_0=72.2_{-4.1}^{+3.6} \mathrm{~km} \mathrm{~s}^{-1} \mathrm{Mpc}^{-1}$. Compared with the results from six gravitationally lensed quasars with measured time delays $H_0=73.3_{-1.8}^{+1.7} \mathrm{~km} \mathrm{~s}^{-1} \mathrm{Mpc}^{-1}$ \cite{Wong:2019kwg}, the central value of $H_0$ decreases a little bit.

\begin{figure}
	\begin{center}
		\includegraphics[width=8cm]{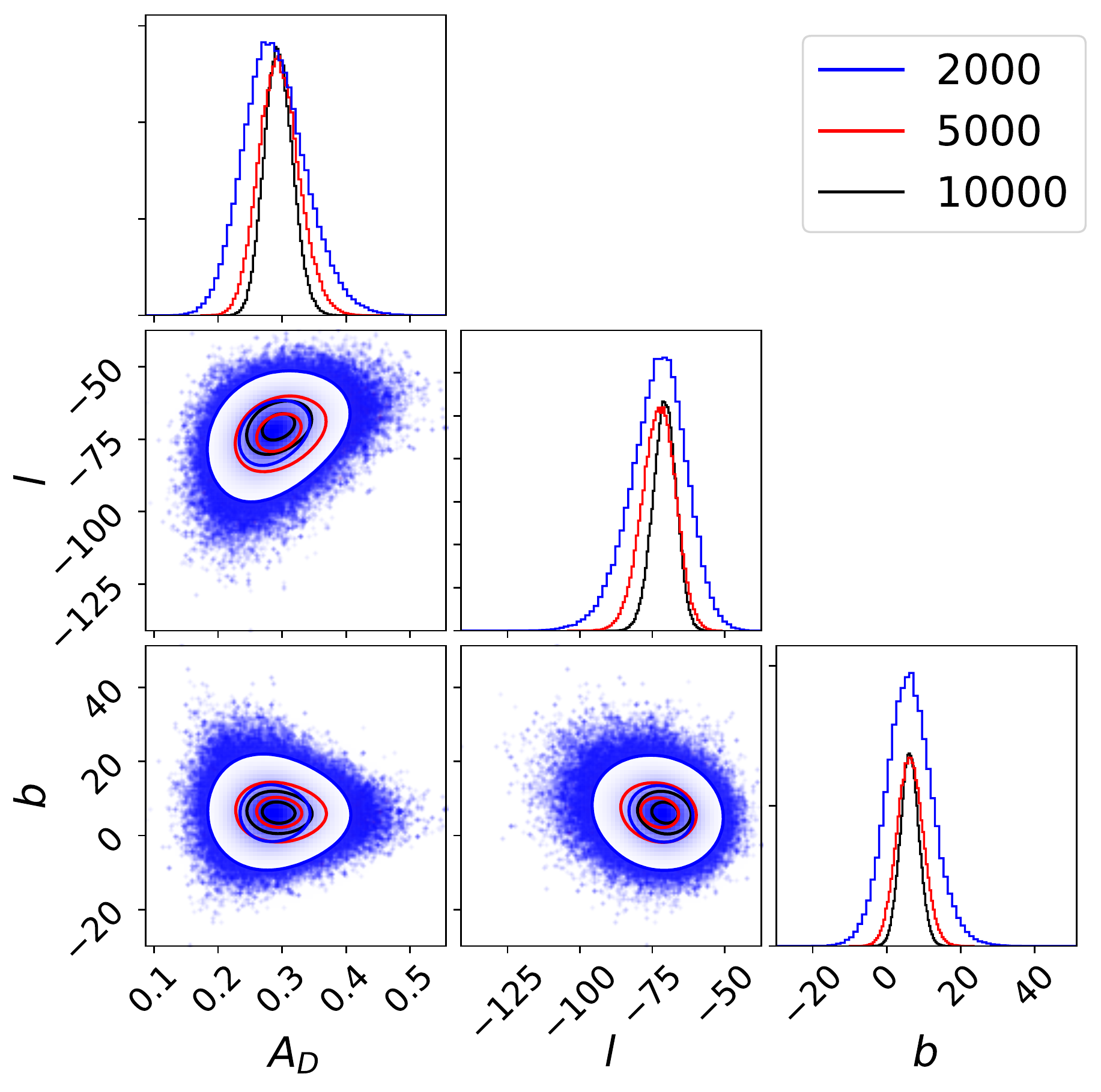}
		\caption{The marginalized posterior distribution of the parameters of dipole with the simulated X-ray and UV fluxes of quasars. The horizontal and vertical solid black lines denote the maximum of 1-dimensional marginalized posteriors. The blue, red, dark lines denote 2000, 5000 and, 10000 simulations, respectively.}
		\label{fig:Sim_result}
	\end{center}
\end{figure}

Finally, we forecast the future constraints on the dipole parameters in the Finslerian cosmological model with the X-ray and UV fluxes of quasars. We assume Finslerian cosmological model with $\Omega_{\mathrm{m}}=0.509$, $A_D=0.302$ and $(l,b)=(288.92^{\circ},6.10^{\circ})$, which are given by the the X-ray and UV fluxes of 808 quasars. In the simulation, the positions of the 808 quasars are unchanged and the redshifts of simulated quasars are generated from the redshift distribution of the 808 quasars. We replace the X-ray fluxes of \textit{i}th simulated quasar with a random number generated from the Gaussian distribution $G(F_{X}^{\mathrm{obs}},\sigma_{F_{X}^{\mathrm{obs}}})$, where the $F_{X}^{\mathrm{obs}}$ is the X-ray fluxes of observed quasars at the same position with the \textit{i}th simulated quasar and $\sigma_{F_{X}^{\mathrm{obs}}}$ is the error of the observed flux $F_{X, i}^{\mathrm{obs}}$. We construct 2000, 5000, and 10000 simulations for comparison, and the results of the dipole parameters from the simulated dataset are shown in Fig. \ref{fig:Sim_result} and summarized in Table \ref{table:S_data}. In Fig. \ref{fig:Sim_result}, the blue, red, and dark lines denote 2000, 5000, and 10000 simulations, respectively. As can be seen, the inferred errors of the dipole parameters get smaller as the number of simulations increases. For the dipole amplitude $A_D$, the inferred error of lower and upper limits reduce to $12.76\%$ and $15.86\%$. For the direction parameters $l$ and $b$, the precisions increase by about 50\% when the 2000 simulations are compared with the 10000 simulations. Our results show that the X-ray and UV fluxes of quasars have a promising future as a probe of the Finslerian cosmological model.

\begin{table}
	\caption{The 68\% confidence level constraints on the parameters in the Finslerian cosmological model with the simulated dataset.}
	\setlength{\tabcolsep}{1.8mm}{
		\begin{tabular}{cccc}
			\hline Simulations & 2000 & 5000 & 10000 \\
			\hline
			$A_D$ & $0.280_{ -0.078}^{ +0.105}$ & $0.289_{ -0.054}^{ +0.065}$ & $ 0.290_{ -0.037}^{ +0.046}$ \\
			$l$ & $288.35_{~ -19.92^{\circ}}^{^{\circ}  +16.15^{\circ}}$ & $289.70_{~ -11.78^{\circ}}^{^{\circ}  +9.81^{\circ}}$ & $289.00_{~ -7.57^{\circ}}^{^{\circ} +7.73^{\circ}}$ \\
			$b$ & $6.66_{~ -12.17^{\circ}}^{^{\circ} +11.45^{\circ}}$ & $5.81_{~ -6.47^{\circ}}^{^{\circ} +7.36^{\circ}}$ & $6.40_{~ -4.87^{\circ}}^{^{\circ} +4.64^{\circ}}$ \\
			\hline
	\end{tabular}}
	\label{table:S_data}
\end{table}
\section{Discussions and conclusions}\label{sec:DC}
In this paper, we tested the anisotropy in the Finslerian cosmological model with the X-ray and UV fluxes of quasars. The dipole anisotropy is well-constrained by the X-ray and UV fluxes of quasars. The dipole amplitude is $A_D=0.302_{ -0.124}^{ +0.185}$ and the dipole direction points towards $(l, b) = ( 288.92_{~ -28.80^{\circ}}^{^{\circ}+23.74^{\circ}}, 6.10_{~ -16.40^{\circ}}^{^{\circ} +16.55^{\circ}} )$. Interestingly, we found that the dipole direction from the X-ray and UV fluxes of quasars is very close to the dipole direction $(l, b) = ( 291.60_{~ -92.96^{\circ}}^{^{\circ} +248.10^{\circ}}, 16.20_{~ -78.73^{\circ}}^{^{\circ} +73.80^{\circ}} )$ given by the JLA in the Finslerian cosmological model. The angular difference between the two dipole directions is only $10.44^{\circ}$. We also found that the dipole direction given by the X-ray and UV fluxes of quasars is not far away from the dipole directions given by SNe Ia in the dipole-modulated $\Lambda$CDM model and the angular difference is around $30^{\circ}$. We thought the dipole direction from the X-ray and UV fluxes of quasars close to the ones from the SNe Ia sample, especially the JLA compilation, may hint that there could exist an underlying relation. We combined the Pantheon sample and JLA compilation with quasars to constrain the Finslerian cosmological model and the results are similar to the ones given by the Pantheon sample in the Finslerian cosmological model \cite{Chang:2019utc}. We also investigated the Hubble constant $H_0$ in the Finslerian cosmological model by combining the X-ray and UV fluxes of quasars with six gravitationally lensed quasars. We found a slightly smaller value of Hubble constant $H_0=72.2_{-4.1}^{+3.6} \mathrm{~km} \mathrm{~s}^{-1} \mathrm{Mpc}^{-1}$ than the value $H_0=73.3_{-1.8}^{+1.7} \mathrm{~km} \mathrm{~s}^{-1} \mathrm{Mpc}^{-1}$ given by the six gravitationally lensed quasars. At last, we forecasted the future constraints on the dipole parameters with the X-ray and UV fluxes of quasars. We constructed 2000, 5000, and 10000 simulations and found that the precisions of the parameters related to anisotropy have a significant improvement as the number of simulations increases. Our results show that the X-ray and UV fluxes of quasars have a promising future as a probe of anisotropy in the Finsler spacetime.
\section*{Acknowledgments}\noindent
We thank Yong Zhou for helpful discussions. J.-Q.Xia is supported by the National Science Foundation of China under grants No. U1931202, 11633001, and 11690023; the National Key R\&D Program of China No. 2017YFA0402600.
%
%
%
\bibliographystyle{prsty}
\bibliography{ref}
%
%
%

\end{document}